\documentclass[preprint,aps,showpacs,showkeys,dvips]{revtex4}

\usepackage{graphicx}% Include figure files
\usepackage{dcolumn}% Align table columns on decimal point
\usepackage{bm}% bold math
\usepackage{subfigure}

\begin{document}

\title{Possible impact of the fourth generation quarks on production of a charged Higgs boson at the LHC}

\author{R. \c{C}ift\c{c}i}
\email{rena.ciftci@gmail.com}
\affiliation{Dept. of Eng. of Physics, Faculty of Eng., Ankara University, 06100 Tandogan,
Ankara, Turkey}

\author{ A. K. \c{C}ift\c{c}i}
\email{ciftci@science.ankara.edu.tr}
\affiliation{Physics Department, Faculty of Sciences, Ankara University, 06100 Tandogan,
Ankara, Turkey}

\author{S. Sultansoy}
\email{ssultansoy@etu.edu.tr}
\affiliation{Physics Section, Faculty of Sciences and Arts, TOBB University of Economics and Technology,
Ankara, Turkey}
\altaffiliation[Also at ]{Institute of Physics, Academy of Sciences, H. Cavid Avenue 33, 
Baku, Azerbaijan}

\date{\today}

\begin{abstract}
We investigate the impact of the fourth generation quarks on production and decays of the charged Higgs boson at CERN Large Hadron Collider (LHC) with triple $b$-tagging. The signal is the process $gg\rightarrow \bar{u_4}u_4$, followed by $\bar{u_4}\rightarrow W^{-} \bar{b}$ and $u_4\rightarrow h^{+} b$ decays with subsequent $h^{+}\rightarrow t \bar{b}$ and corresponding hermitic conjugates. It is shown that, if $m_{u_{4}}= 400$ GeV, considered process will provide unique opportunity to discover charged Higgs boson with mass range of $200$ to $350$ GeV at the first years of the LHC run.
\end{abstract}
\maketitle
\section{Introduction}	

It is known that two-Higgs doublet model (2HDM), in general, and minimal supersymmetric extension of the standard model (MSSM), in particular, predict the existence of a charged scalar particle as well as two neutral scalar particles in addition to the standard model (SM) Higgs boson {[}1{]}. Experimental observations of these particles could be indirect indication of SUSY. Experiments at LEPII limit the mass of a charged Higgs boson from below as $79.2$ GeV {[}2{]}. The Tevatron CDF excludes masses of a charged Higgs boson below $105$ and $130$ GeV for $\tan\beta=1$ and $\tan\beta=40$, respectively, by searching $t\rightarrow h^{\pm} b$ decays {[}3{]}. Obviously, higher energy reach of the Large Hadron Collider (LHC) will give opportunity to search charged Higgs boson in wider mass region. The production of the charged Higgs boson at the LHC for three SM generation case is considered in a number of papers {[}4-9{]}.

On the other hand, flavor democracy, which is quite natural in the SM framework, predicts the existence of the fourth generation (see review {[}10{]} and references therein). The masses of the fourth generation quarks and charged leptons are expected to be almost degenerate with preferable range of $m_4=300\div500$ GeV. Obviously, the fourth generation quarks in this mass region will be observed at the first few years of the LHC data taking {[}11-14{]}. Naturally, as the Tevatron searches $h^{\pm}$ in $t$-quark decays, the LHC may do the same in $u_4$-quark decays.   

In this paper, we investigate the impact of the fourth generation quarks on production and decays of the charged Higgs boson of 2HDM at the LHC. In Section II, the lagrangian describing decays of the charged Higgs is presented and the branching ratios of decays of the fourth SM generation up quark and charged Higgs boson are evaluated. The production of the charged Higgs boson at the LHC via gluon-gluon fusion process $gg\rightarrow \bar{u_4}u_4$, followed by $\bar{u_4}\rightarrow W^{-} \bar{b}$ and $u_4\rightarrow h^{+} b$ decays with subsequent $h^{+}\rightarrow t \bar{b}$, as well as the SM background is studied in Section III. The statistical significance of the charged Higgs boson signal at the LHC is estimated assuming three $b$-quark jets to be tagged. Finally, concluding remarks are made in Section IV.

\begin{figure}
\subfigure[]{\includegraphics[width=12.0cm]{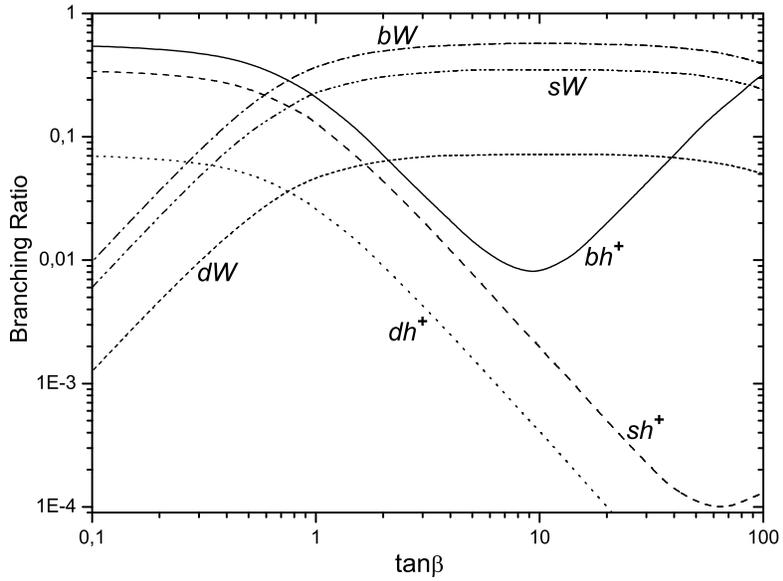}}
\subfigure[]{\includegraphics[width=12.0cm]{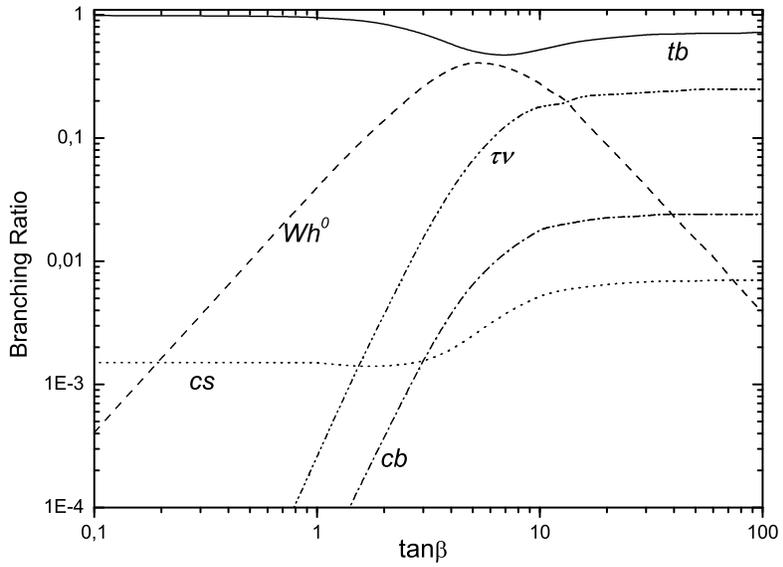}}
\caption{Branching ratios for different decay channels of the (a) fourth generation up quark and (b) charged Higgs boson  as a function of the $\tan\beta$ in the 2HDM. Following mass values are used: $m_{u_{4}}= 400$ GeV, $m_{h^{\pm}}= 200$ GeV and $m_{h^{0}}= 115$ GeV. \protect\label{fig1}}
\end{figure}

\begin{figure}
\subfigure[]{\includegraphics[width=12.0cm]{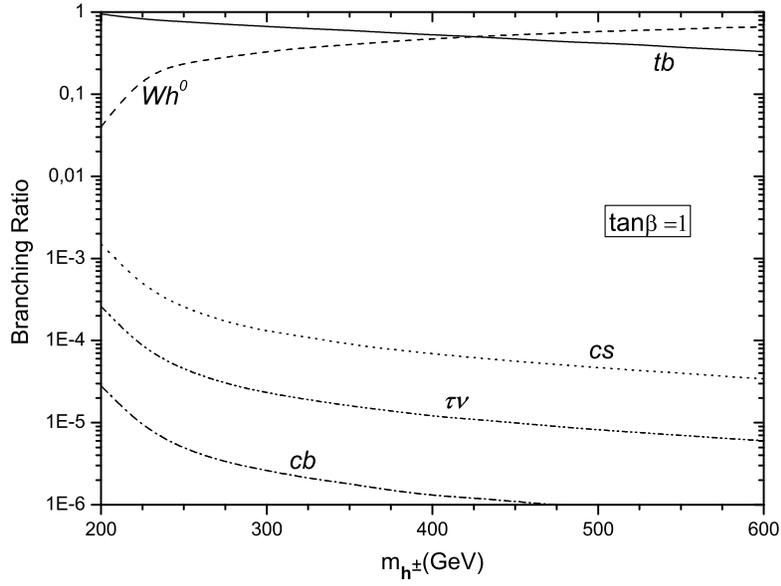}}
\subfigure[]{\includegraphics[width=12.0cm]{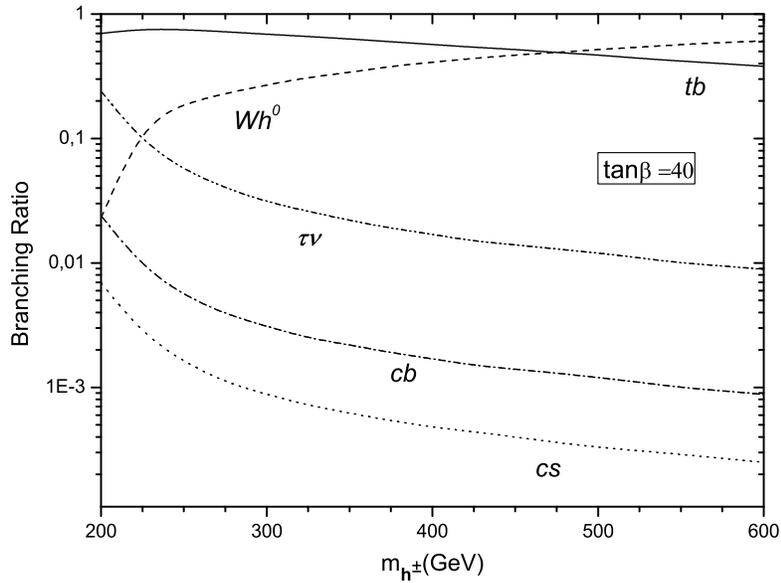}}
\caption{Branching ratios for decays of the charged Higgs boson as a function of its mass for different values of $\tan\beta$: (a) $\tan\beta=1$ and (b) $\tan\beta=40$. \protect\label{fig2}}
\end{figure}

\section{Charged Higgs boson decays}

Interactions involved charged Higgs boson can be described as below {[}6{]}:
\begin{equation}
{\cal L}=\frac{g}{\sqrt{2}} h^{+}\left(\cot{\beta}\; m_{u_i} \overline u_{i} d_{iL}+ \tan{\beta}\; m_{d_i} \overline u_{i} d_{iR}+ \tan{\beta}\; m_{l_i} \overline \nu_{i} l_{iR}\right)+h.c. \hspace{2mm} ,  \hspace{3mm} 
\end{equation}
where $\textit{i}=1,2,3,4$ denotes the generation index and $\tan\beta$ is defined as ratio of the two Higgs doublets vacuum expectation values. Applying the flavor democracy to three generation MSSM result in $\tan\beta = {m_t}/{m_b}\approx 40$ {[}10{]}, whereas $\tan\beta \approx 1$ is preferable in four generation case. The Cabibbo-Kobayashi-Maskawa (CKM) matrix elements are not shown in Eq. (1). In numerical calculations, we use CKM mixings given in Ref. {[}15{]}.

\begin{figure}[b]
\includegraphics[width=12.0cm]{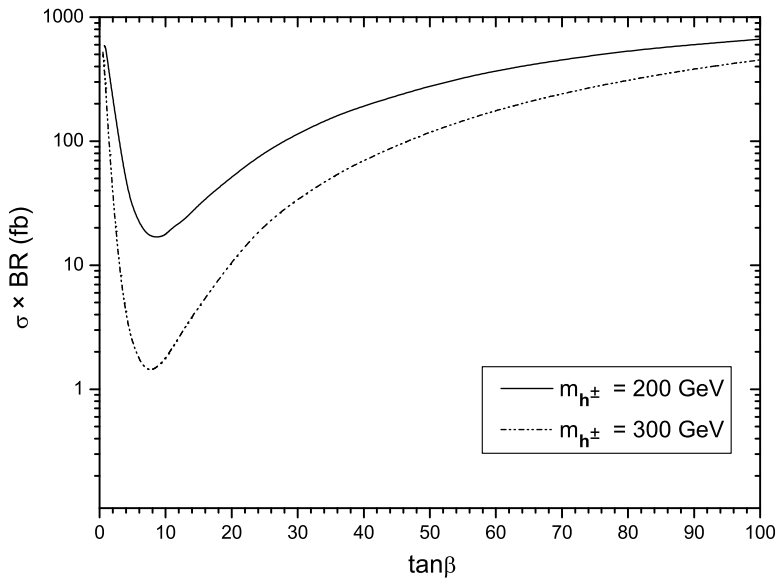}
\caption{$\sigma \times BR$ of the process $gg\rightarrow \bar{u_4}u_4\rightarrow W^{-} \bar{b} b h^{+}\rightarrow W^{-} \bar{b} b t \bar{b}\rightarrow W^{-} \bar{b} b \bar{b} b W^{+}\rightarrow \bar{b} b \bar{b} b l\nu \bar{q} q'$ for a charged Higgs boson with $200$ and $300$ GeV masses as a function of $\tan\beta$ \label{fig3}.}
\end{figure}

In order to compute decay widths of the charged Higgs boson, above lagrangian has been implemented into the CompHEP {[}16{]}. The decay branching ratios of the fourth generation up quark with mass of $400$ GeV (which is used at the rest of the paper) are plotted in Fig. 1a for $m_{h^{\pm}}= 200$ GeV. These plots show that the dominant decay channels of $u_4$ are $bh^{+}$ and $sh^{+}$ at low $\tan\beta$ values; $bW$ and $sW$ decays are dominant at $\tan\beta>1$ region. 

Obtained results for branching ratios of decays of the charged Higgs boson into SM fermions are given in Fig. 1b as a function of $\tan\beta$. The charged Higgs boson dominantly decays to $tb$ for almost all $\tan\beta$ values. Furthermore, Figs. 2a and 2b present the branching ratios of the charged Higgs boson decays as a function of its mass for two different values of $\tan\beta$, 1 and 40, respectively.

\section{Charged Higgs Boson Production at the LHC}

We study the $gg\rightarrow \bar{u_4}u_4\rightarrow W^{-} \bar{b} b h^{+}\rightarrow W^{-} \bar{b} b t \bar{b}\rightarrow W^{-} \bar{b} b \bar{b} b W^{+}$ (and its hermitic conjugate) production process at the LHC, followed by leptonic decay of one $W$ and hadronic decay of the other. The calculated production cross sections with $m_{u_{4}}= 400$ GeV are plotted in Fig. 3 for charged Higgs boson mass values of $200$ and $300$ GeV. CTEQ6L1 parton distribution functions {[}17{]} are used in numerical calculations. The SM background cross sections are computed using MadGraph package {[}18{]}. This background is potentially much larger than the signal. However, in order to extract the charged Higgs boson signal and to suppress the SM background, we impose some kinematic cuts. In addition, we assume that three $b$-quark jets are tagged.

\begin{figure}
\subfigure[]{\includegraphics[width=11.5cm]{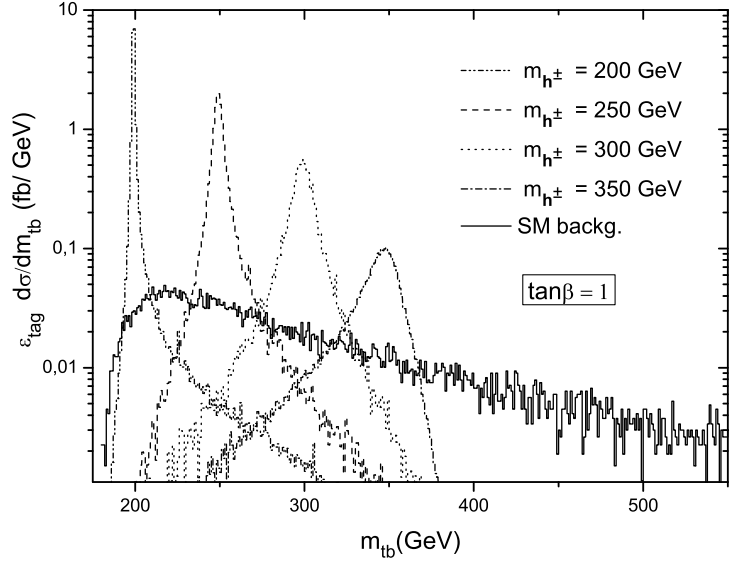}}
\subfigure[]{\includegraphics[width=11.5cm]{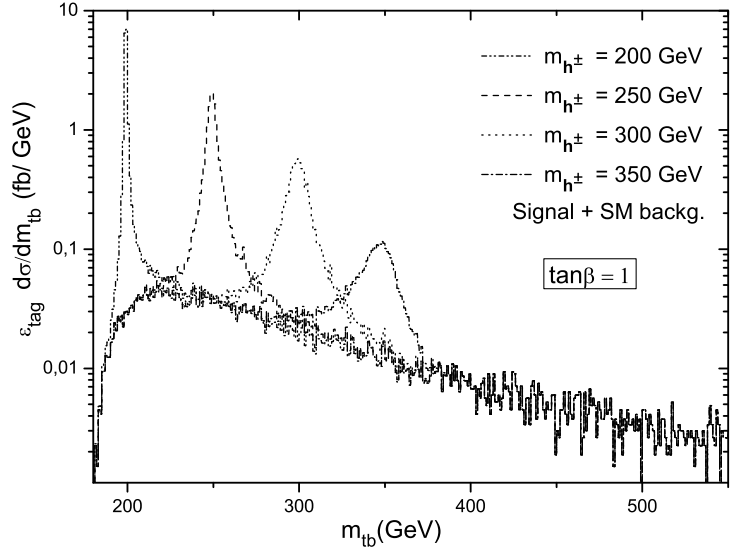}}
\caption{The signal cross section distributions of the reconstructed charged Higgs boson mass for four selected mass values at $\tan\beta\:=\:1$ and corresponding SM background: (a) separated and (b) summed. \protect\label{fig4}}
\end{figure}

\begin{figure}
\subfigure[]{\includegraphics[width=11.5cm]{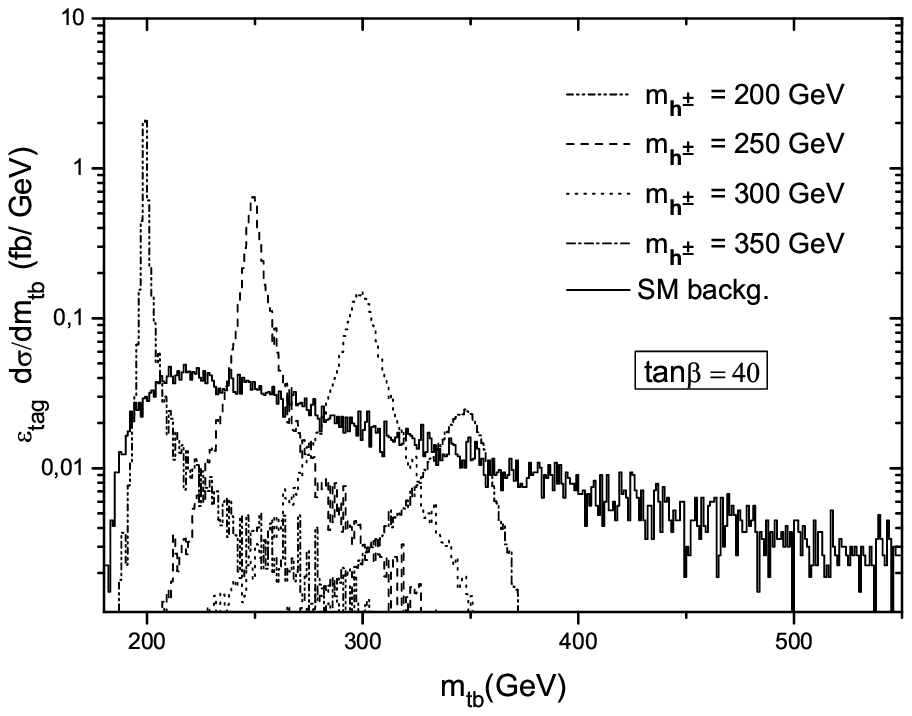}}
\subfigure[]{\includegraphics[width=11.5cm]{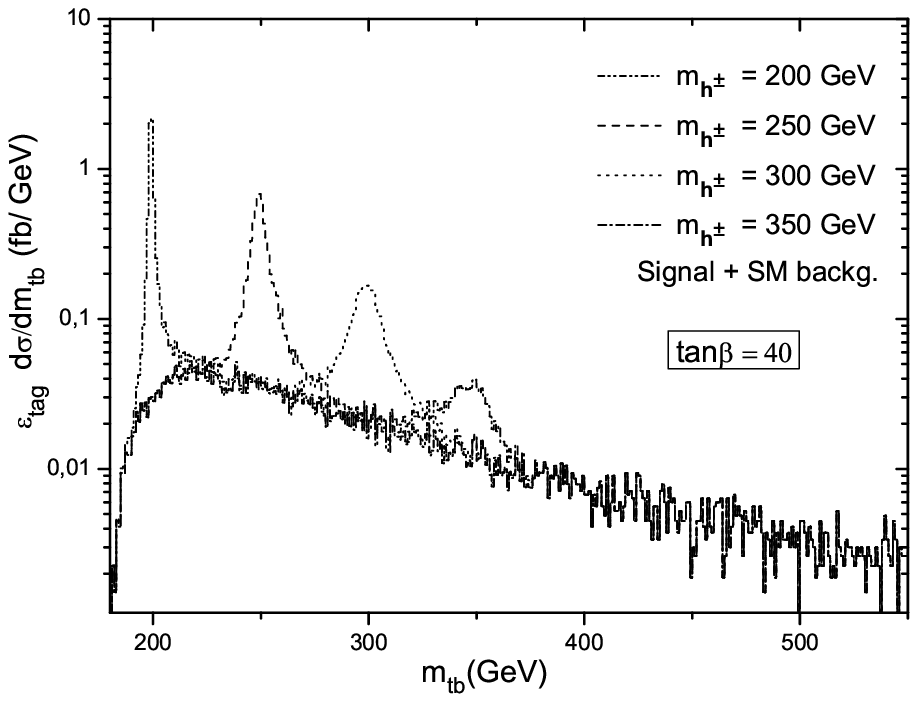}}
\caption{The same as Fig. 5 but for $\tan\beta\:=\:40$. \protect\label{fig5}}
\end{figure}

We choose the following set of selection cuts: $P_T\:>\:20$ GeV for all the jets and the lepton, excluding $P_T\:>\:80$ GeV cut for at least one of $b$-jets; $\left|\eta\right|\:<\:2.5$, where $\eta$ denotes pseudorapidity; a minimum separation of $\Delta R\:=\:\left[\left(\Delta\phi\right)^{2}+\left(\Delta\eta\right)^{2}\right]^{1/2}\:>\:0.4$ ($\phi$ is the azimuthal angle) between the lepton and the jets as well as each pair of jets. The signal and background cross sections are given in Fig. 4 as a function of the reconstructed $tb$ invariant mass. It is drawn for sample values of the charged Higgs boson masses of $200$, $250$, $300$ and $350$ GeV for $\tan\beta\:=\:1$. Here we have included a $b$-tagging efficiency of $\%50$. The signal and SM background cross sections are shown separately in Fig. 4a, while their sum is presented in Fig. 4b. The signal peaks are clearly visible at all selected mass values. The similar plots for $\tan\beta\:=\:40$ are presented in Fig. 5a and 5b.

The number of events - in a window of $40$ GeV around selected $m_{h^{\pm}}$ values - for signal (S), and SM background (B), along with the statistical significance ($S/\sqrt{B}$) for $100$ fb$^{-1}$ and $10$ fb$^{-1}$ of integrated luminosity are presented in Table I and II for $\tan\beta\:=\:1$ and $\tan\beta\:=\:40$, respectively. It is seen that, the mass regions $m_{h^{\pm}}= 200\div350$ GeV for $\tan\beta\:=\:1$ and $m_{h^{\pm}}= 200\div300$ GeV for $\tan\beta\:=\:40$ are covered with more than $5\sigma$ even with low integrated luminosity of $10$ fb$^{-1}$. To compare with three SM generation case, for example, at $\tan\beta\:=\:40$ and $m_{h^{\pm}}= 300$ GeV we obtain the signal significance $9.6\sigma$ with $10$ fb$^{-1}$, whereas $S/\sqrt{B}$ is $6.2$ with $100$ fb$^{-1}$ in three SM generation case {[}7{]}.    

\begin{table}
\caption{The number of charged Higgs boson signal and SM background events and the statistical significance of the signal for $\tan\beta=1$}
\begin{ruledtabular}
\begin{tabular}{ccccccc}
& \multicolumn{2}{c} {$S$} &\multicolumn{2}{c} {$B$}&\multicolumn{2}{c} {$S/\sqrt{B}$}  \\
\cline{2-3}\cline{4-5}\cline{6-7}
$m_{h^{\pm}}\pm 20$ GeV &$100$ fb$^{-1}$ & $10$ fb$^{-1}$  & $100$ fb$^{-1}$ & $10$ fb$^{-1}$& $100$ fb$^{-1}$ & $10$ fb$^{-1}$ \\ 
\colrule

200\hphantom{00} & \hphantom{0}1710 & \hphantom{0} 171 & \hphantom{0}105 & 10.5 &166.9 &52.8\\
250\hphantom{00} & \hphantom{0}1540 & \hphantom{0} 154 & \hphantom{0}134 & 13.4 &133.0 &42.1\\
300\hphantom{00} & \hphantom{0}833  & \hphantom{0} 83.3  & \hphantom{0}75 & 7.5 &96.3  &30.4\\
350\hphantom{00} & \hphantom{0}215  & \hphantom{0} 21.5  & \hphantom{0}47  & 4.7  &31.3 &9.9\\
\end{tabular}
\end{ruledtabular}
\end{table}

\begin{table}
\caption{The same as Table I but for $\tan\beta=40$}
\begin{ruledtabular}
\begin{tabular}{ccccccc}
& \multicolumn{2}{c} {$S$} &\multicolumn{2}{c} {$B$}&\multicolumn{2}{c} {$S/\sqrt{B}$}  \\
\cline{2-3}\cline{4-5}\cline{6-7}
$m_{H^{\pm}}\pm 20$ GeV &$100$ fb$^{-1}$ & $10$ fb$^{-1}$  & $100$ fb$^{-1}$ & $10$ fb$^{-1}$& $100$ fb$^{-1}$ & $10$ fb$^{-1}$ \\ 
\colrule

200\hphantom{00} & \hphantom{0}630 & \hphantom{0} 63.0 & \hphantom{0}105 & 10.5 &61.5 &19.4\\
250\hphantom{00} & \hphantom{0}599 & \hphantom{0} 59.9 & \hphantom{0}134 & 13.4 &51.8 &16.4\\
300\hphantom{00} & \hphantom{0}263 & \hphantom{0} 26.3 & \hphantom{0}75 & 7.5 &30.4 &9.6\\
350\hphantom{00} & \hphantom{0}57  & \hphantom{0} 5.7 & \hphantom{0}47  & 4.7  &8.2 &2.6\\
\end{tabular}
\end{ruledtabular}
\end{table}

\section{Conclusion}

Our study show that, the existence of the fourth SM generation provides new channel for charged Higgs boson search at the LHC. If the fourth generation quarks and charged Higgs boson has appropriate masses, this channel will be a discovery mode. More detailed study including higher $m_{u_{4}}$ mass values, as well as, further optimizations of cuts, detector features etc. is ongoing.    

\begin{acknowledgments}

R. Ciftci would like to acknowledge for support from the Scientific and Technical Research Council (TUBITAK) BIDEB-2218 grant. This work was also supported in part by the State Planning Organization (DPT) under grant no DPT-2006K-120470 and in part by the Turkish Atomic Energy Authority (TAEA) under grant no VII-B.04.DPT.1.05.

\end{acknowledgments}

\end{document}